\documentclass[lnicst,sechang,a4paper]{svmultln}
\usepackage{amssymb}
\usepackage{graphicx}

\usepackage{url}
\urldef{\mailsa}\path|{maha.alodeh, symeon.chatzinotas, bjorn.ottersten}@uni.lu|
\usepackage[pdfpagelabels,hypertexnames=false,breaklinks=true,bookmarksopen=true,bookmarksopenlevel=2]{hyperref}

\begin{document}

\mainmatter  

\title{Symbol Based Precoding in The Downlink of Cognitive MISO Channel}

\titlerunning{Symbol Based Precoding in The Downlink of Cognitive MISO Channel}

%
%
\author{Maha Alodeh%
\thanks{This work is supported by Fond National de la Recherche Luxembourg (FNR)
projects,
project Smart Resource Allocation for Satellite Cognitive Radio (SRAT-SCR)  ID:4919957 and Spectrum Management and Interference Mitigation in Cognitive Radio Satellite Networks SeMiGod.}%
\and Symeon Chatzinotas\and Bj\"orn Ottersten}  %
\authorrunning{Symbol Based Precoding in The Downlink of Cognitive MISO Channel}

\institute{Interdisciplinary Centre for Security, Reliability and Trust, University of Luxembourg\\ 4, Alphonse Weicker, 2721 Luxembourg\\
\mailsa
}

%
%

\toctitle{Symbol Based Precoding in The Downlink of Cognitive MISO Channel}
\tocauthor{Authors' Instructions}
\maketitle
\begin{abstract}
This paper proposes symbol level precoding in the 
downlink of a MISO cognitive system. The new scheme tries to jointly utilize the data and channel information to design a precoding that minimizes the transmit power at a cognitive
base station (CBS); without violating the interference temperature constraint
imposed by the primary system. In this framework, the data information is
handled at symbol level which enables the characterization  the intra-user interference among
the cognitive users as an additional source of useful energy that should
be exploited. A relation between the constructive multiuser transmissions
and physical-layer multicast system is established.
Extensive simulations are performed to validate the proposed technique and compare it with conventional techniques.
\end{abstract}
\vspace{-0.4cm}
\section{Introduction}
The combination of the spectrum scarcity and congestion has motivated researchers to propose more innovative techniques to tackle these challenges. Fixed spectrum allocation techniques assign certain bands to certain applications, which may no longer efficiently used \cite{federal}.
Solving the problem would require changing the regulations which is a complicated and lengthy
procedure. With that in mind, the paradigm of cognitive radios has been proposed as a promising agile technology that can revolutionize the future of telecommunication by ``breaking the gridlock of the wireless spectrum" \cite{goldsmith}. The key idea of their implementation is to allow opportunistic transmissions to share the wireless medium. Thus, two initial hierarchical levels have been defined: primary level and cognitive level (the users within each level are called primary users (PU) and cognitive users (CU) respectively). The interaction between these two levels is determined by the agility of the cognitive level and the predefined constraints imposed by the primary level \cite{Haykin}. Overlay,
underlay and interweave are three general implementations which regulate the
coexistence terms of both systems. The first two implementations allow simultaneous transmissions, which leads to better spectrum utilization in comparison to the last one, which allocates the spectrum to the cognitive system by detecting the absence of the primary one \cite{shree}.\smallskip
      
The form of integration in this work is defined by cooperation between
the two levels in the cognitive interference channel.  The cooperation can aid the primary network to satisfy the quality of service (QoS) or enhance
the rate of its
own users by backhauling its data through the cognitive system\cite{lv_o}-\cite{gan_t}; CBSs can operate as relays for primary messages and as regular base stations to serve their cognitive users. The cognitive system benefits by providing a service to its users. This kind of cognitive implementation fits with practical overlay cognitive definition, as the PU is being served from
both the PS and the CBSs by performing relaying between them to make primary data accessible by the CBS. Sometimes the primary symbols are not  available to the cognitive
system,  
as a result the cognitive system needs to take the sufficient precautions
to protect the primary system from the interference created by its own transmissions.
It should be noted that we assume that the CBSs are equipped with multiple antennas to handle multi-user transmissions, and to enable interference
mitigation.\smallskip

The conventional look at interference can  be shifted from a degradation factor
into a favorable one if we handle the transmitted data frame at symbol level.
At this level, the interference can be classified into: constructive and destructive  ones. This classification 
is initially proposed in \cite{Christos-1}; instead of fully inverting the
channel to grant zero interference among the spatial streams, the proposed
precoding suggests keeping the constructive interference while removing
the destructive part by partial channel inversion.
This technique is proven to outperform the traditional
zero forcing precoding. A more advanced technique is proposed
in  \cite{Christos}, where an interference rotation  is examined to make the interference constructive for all users.  Moreover, a modified maximum
ratio transmissions technique that performs unitary rotations to create constructive
interference among the interfering multiuser streams is proposed \cite{maha}.
Furthermore, a connection between symbol based constructive interference precoding and PHY multicast is established
in \cite{maha}-\cite{maha_TSP}.

In this work, we utilize the symbol level precoding in underlay MISO cognitive
radio scenarios. We shape the interference between the cognitive users to
provide constructive characteristics without violating the interference temperature constraints on the primary
receivers.  
\vspace{-0.2cm}
\section{System and Signals model}
\hspace{-2cm}\begin{figure*}[t]
\vspace{-0.5cm}
\begin{tabular}[t]{c}
\begin{minipage}{12 cm}
\begin{center}
\vspace{-1cm}
\hspace{-0.9cm}\includegraphics[scale=0.5]{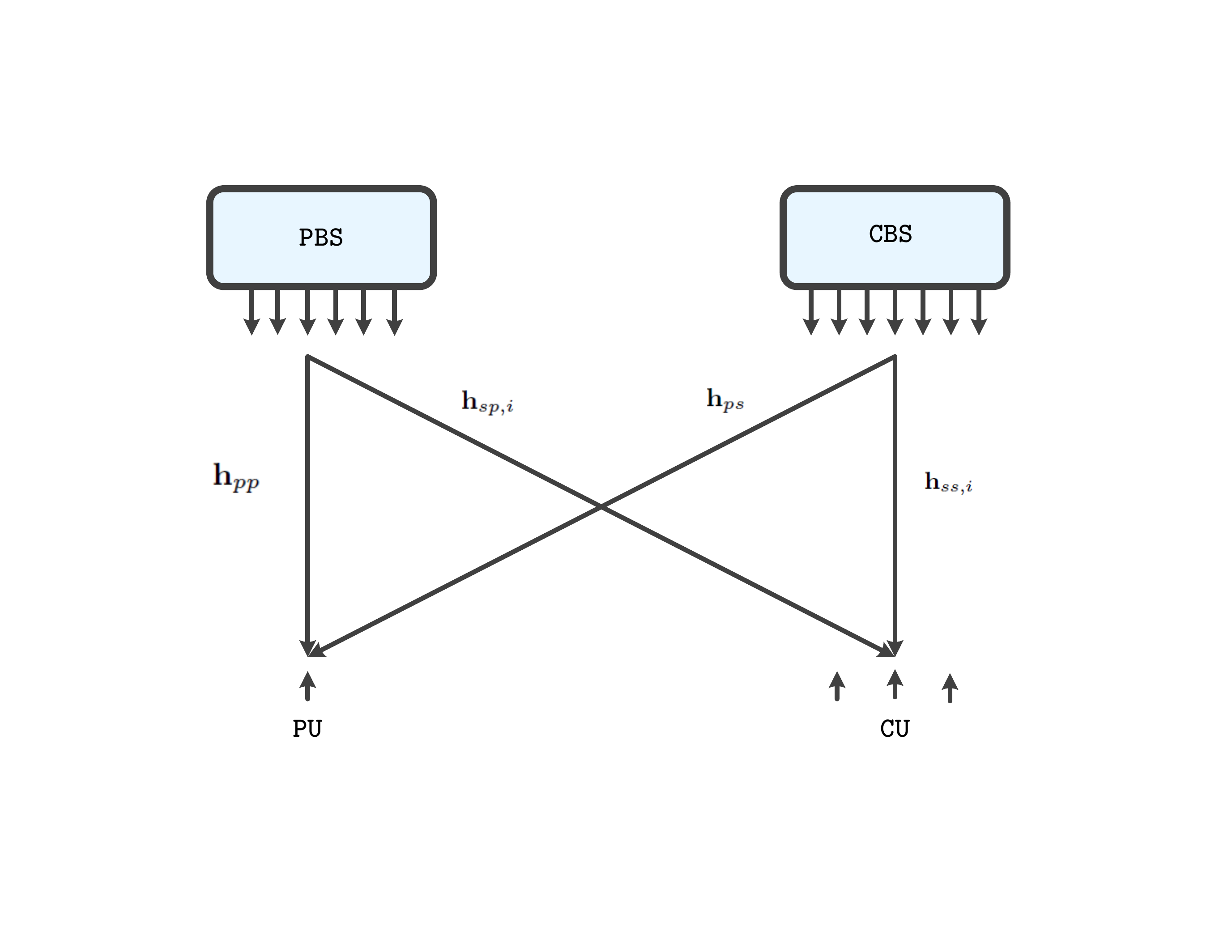}
\vspace{-2.9cm}
\caption{\label{model}\textit{\label{fig1}\small System model 
}}
\end{center}
\end{minipage}\\
\hline
\end{tabular}
\end{figure*}

We consider a cognitive radio network which shares the spectrum
resource with a primary network in the underlay mode as fig. (\ref{model}).
The primary network consists of a primary base station (PBS), equipped with
$N_p$ antennas, serving
 a single primary user. The
cognitive network has a single CBS,
equipped with $M$ antennas, serving $K$ CU. Each CU
is equipped with a single antenna. Throughout this
paper, we consider that $K\leq M-1$ and that the
primary user is equipped with a single antenna. Due to the
sharing of the same frequency band, the received signal at
the primary user is interfered by the signals transmitted from
CBS. Similarly, the received signals at the CUs are
interfered by the signal transmitted from the PBS.\smallskip

Assume that in one time slot, a block of information
symbols $\mathbf{d} = [d_1, d_2,..., d_K]^T$ are sent from the CBS in
which $d_k$, $k = 1, ... ,K$ is the desired signal for user $k$.
We assume that $\mathbf{d}$ contains uncorrelated unit-power M-PSK entries.
With a proper beamforming (which will
be specified later), the transmit signal is given by
\begin{eqnarray}
\vspace{-0.1cm}
\mathbf{x}=\mathbf{W}\mathbf{d}
\end{eqnarray}
where $\mathbf{W}=[\mathbf{w}_1, \mathbf{w}_2, ...,\mathbf{w}_K]$ denotes the transmit
precoding matrix for the cognitive system while $\mathbf{w}_k\in\mathbb{C}^{M\times
1}$ denotes the beamforming vector for $k^{th}$ CU. The received signal at the $k^{th}$
user, denoted by $y_{s,k}$, is given by
\begin{eqnarray}
\hspace{-0.9cm}y_{s,k}=\mathbf{h}_{ss,k}\mathbf{w}_k d_k+\sum_{j\in K,j\neq k}\mathbf{h}_{ss,k}\mathbf{w}_j d_j+\mathbf{h}_{sp,k}\mathbf{g}^pd_p
+n_k,
\end{eqnarray}
and the received signal at PU's receiver is given by

\begin{eqnarray}
\hspace{-0.9cm}y_{p}=\mathbf{h}_{pp}\mathbf{g} d_p+\sum_{j\in K}\mathbf{h}_{ps}\mathbf{w}_j d_j+n
\end{eqnarray}
where $\mathbf{h}_{ss,k}\in\mathbb{C}^{1\times M}$ and $\mathbf{h}_{sp,k}\in\mathbb{C}^{1\times
N_p}$ are the channels between the CBS and the PBS respectively and  the $k^{th}$ CU. 
 While $\mathbf{h}_{pp}$ and $\mathbf{h}_{ps}$ denote the channel between the
 PBS and PU, CBS and PU respectively. The transmitted power
of the primary user is denoted by $p_p$, $\mathbf{g}\in\mathbb{C}^{N_p\times 1}$ represents the precoding vector
used by the PBS, and $d_p$
represents the transmitted symbol from the PBS and it is not available at CBS.
Finally, ${n}_k\sim\mathcal{CN}(0,\sigma^2)$ and $n\sim \mathcal{CN}(0,\sigma^2)$ are additive i.i.d.
complex Gaussian noise with zero mean and variance $\sigma^2_k$ at the $k^{th}$
CU and PU respectively. The channel state information (CSI) $\mathbf{h}_{ps}$ and $\mathbf{h}_{ss,j}$ are available at the CBS.

\vspace{-0.3cm}
\section{Constructive interference Definition}
The interference is a random deviation which can move the desired constellation point in any direction. To address this problem, the power of the interference has been used in the past to regulate its effect on the desired signal point. 
The interference among the multiuser spatial streams
leads to deviation of the received symbols outside of their detection region. However,
in symbol level precoding (e.g. M-PSK) this interference pushes the received symbols further into the correct detection region and, as a consequence it enhances the system performance. Therefore, the interference can
be classified into constructive or destructive based on whether it facilitates or deteriorates the correct detection of the received symbol. For BPSK and QPSK scenarios, a detailed classification of interference is discussed thoroughly in \cite{Christos-1}. In this section, we describe the required conditions to have constructive interference for any M-PSK modulation.
\vspace{-0.1cm}
\subsection{Constructive Interference Definition}

Assuming both the data symbols and CSI are available at the CBS, the unit-power
 created interference from the $k^{th}$ data stream on $j^{th}$ user can be formulated as:
\vspace{-0.1cm}
\begin{equation}
\rho_{jk}=\frac{\mathbf{h}_{ss,j}\mathbf{w}_k}{\|\mathbf{h}_{ss,j}\|\|\mathbf{w}_k\|}.
\end{equation}
Since the adopted modulations are M-PSK ones, a definition for
constructive interference can be stated as\smallskip
\begin{newtheorem}{lem}{Lemma}
\end{newtheorem}

\begin{lem}

\label{lemma}\cite{maha_TSP}
For any M-PSK modulated symbol $d_k$, it is said to receive constructive
interference from another simultaneously transmitted symbol $d_j$ which is
associated with $\mathbf{w}_j$ if and only if the following inequalities hold   
\begin{equation}\nonumber
\label{one}
\angle{d_j}-\frac{\pi}{M}\leq \arctan\Bigg(\frac{\mathcal{I}\{\rho_{jk}d_{k}\}}{\mathcal{R}\{\rho_{jk}d_{k}\}}\Bigg)\leq \angle{d_j}+\frac{\pi}{M},
\end{equation}
\begin{equation}\nonumber
\label{two}
\mathcal{R}\{{d_k}\}.\mathcal{R}\{\rho_{jk}
d_{j}\}>0, \mathcal{I}\{{d_k}\}.\mathcal{I}\{\rho_{jk}d_{j}\}>0.\\
\end{equation}
\end{lem}\smallskip
 \vspace{0.1cm}

\begin{newtheorem}{cor}{Corollary}
\end{newtheorem}

\begin{cor}\cite{maha_TSP}
The constructive interference is mutual.
If the symbol $d_j$ constructively interferes with $d_k$, then
the interference from transmitting  the symbol $d_k$ 
is constructive to $d_j$.
\end{cor}
\vspace{-0.05cm}
\section{Constructive interference exploitation}

\vspace{-0.05cm}
\subsection{Relaxed Interference Constraint}
The precoding aims at exploiting the constructive interference among the
cognitive users without violating the interference temperature constraint
imposed by the primary system $\mathcal{I}_{th}$. The optimization can be
formulated as

\begin{eqnarray}
\label{pow}
\hspace{-0.8cm}\mathbf{w}_1, ...,\mathbf{w}_K&=&\mathop{\arg\min}\limits_{\mathbf{w}_1,...,\mathbf{w}_K}\quad \|\sum^K_{k=1}\mathbf{w}_kd_k\|^2\\\nonumber
\hspace{-0.1cm}s.t.&\mathcal{C}_1&:\angle(\mathbf{h}_{ss,j}\sum^K_{k=1}\mathbf{w}_k
d_k)=\angle(d_j),
\forall j\in K\\\nonumber
&\mathcal{C}_2&:\frac{\|\mathbf{h}_{ss,j}\sum^K_{k=1}\mathbf{w}_kd_k\|^2}{\sigma^2+\|\mathbf{h}_{sp,j}\mathbf{g}\|^2}\geq\zeta_j\quad
, \forall j\in K\\\nonumber
&\mathcal{C}_3&:\|\mathbf{h}_{ps}\sum_{k=1}\mathbf{w}_kd_k\|^2\leq \mathcal{I}_{th}
\end{eqnarray}
The first two sets of constraints $\mathcal{C}_{1}$ and $\mathcal{C}_{2}$
grant the reception of the data symbols with certain SNR level $\zeta_j$. The third
constraint $\mathcal{C}_3$ is to protect the PU from the cognitive systems
transmissions. In order to solve (\ref{pow}), we formulate it by using $\mathbf{x}=\sum^K_{k=1}\mathbf{w}_kd_k$
as the following 
 \begin{eqnarray}
 \hspace{-0.7cm}\mathbf{x}=&\arg\mathop{\min}\limits_{\mathbf{x}}&\quad \|\mathbf{x}\|^2\\\nonumber
 \hspace{-0.1cm}s.t.&\mathcal{C}_1:&\frac{\mathbf{h}_{ss,j}\mathbf{x}+\mathbf{x}^H\mathbf{h}^H_{ss,j}}{2}=\sqrt{\psi_i\zeta_{j}}\mathcal{R}\{d_j\}\quad
 , \forall j\in K\\\nonumber
 &\mathcal{C}_2:&\frac{\mathbf{h}_{ss,j}\mathbf{x}-\mathbf{x}^H\mathbf{h}^H_{ss,j}}{2i}=\sqrt{\psi_i\zeta_{j}}\mathcal{I}\{d_j\}\quad , \forall j\in K\\\nonumber
&\mathcal{C}_3:&\|\mathbf{h}_{ps}\mathbf{x}\|^2\leq \mathcal{I}_{th}.
\end{eqnarray} 
where $\psi_j=\sigma^2+\|\mathbf{h}_{sp,j}\mathbf{g}\|^2$. To solve the problem, the corresponding Lagrange function can be expressed as
 \begin{eqnarray}
 \label{SUopt}
\hspace{-0.8cm}\mathcal{L}(\mathbf{x})&=&\|\mathbf{x}\|^2\\\nonumber
&+&\sum_j{\mu_j}\bigg(-0.5i\small(\mathbf{h}_{ss,j}\mathbf{x}-\mathbf{x}^H\mathbf{h}^H_{ss,j}\small)-\sqrt{\psi_j\zeta_{j}}\mathcal{I}\{d_j\}\bigg)\\\nonumber
&+&\sum_j{\alpha_j}\bigg(0.5\small(\mathbf{h}_{ss,j}\mathbf{x}+\mathbf{x}^H\mathbf{h}^H_{ss,j}\small)
-\sqrt{\psi_j\zeta_{j}}\mathcal{R}\{d_j\}\bigg)\\\nonumber
&+&\lambda\bigg(\mathbf{x}^H\mathbf{h}^H_{ps}\mathbf{h}_{ps}\mathbf{x}-\mathcal{I}_{th}\bigg).\\\nonumber
\end{eqnarray}The KKT conditions can be derived as
\begin{eqnarray}\nonumber
\label{kkt1}
&\hspace{-1.2cm}\frac{d\mathcal{L}(\mathbf{x},\mu_j,\alpha_j,\lambda)}{d\mathbf{x}^*}&=\mathbf{x}+\sum
0.5i\mu_j\mathbf{h}^H_{ss,j}+\sum_j 0.5\alpha_j\mathbf{h}^H_{ss,j}+\lambda\mathbf{h}^H_{ps}\mathbf{h}_{ps}\mathbf{x}\\\nonumber
\label{kkt2}
&\hspace{-1.2cm}\frac{d\mathcal{L}(\mathbf{x},\mu_j,\alpha_j,\lambda)}{d\mu_j}&=-0.5i\small(\mathbf{h}_{ss,j}\mathbf{x}-\mathbf{x}^H\mathbf{h}^H_{ss,j}\small)-\sqrt{\psi_j\zeta_{j}}\mathcal{I}\{d_j\},
\forall
j\in K\\\nonumber
\label{kkt3}
&\hspace{-1.2cm}\frac{d\mathcal{L}(\mathbf{x},\mu_j,\alpha_j,\lambda)}{d\alpha_j}&=0.5\small(\mathbf{h}_{ss,j}\mathbf{x}+\mathbf{x}^H\mathbf{h}^H_{ss,j}\small)
-\sqrt{\psi_j\zeta_{j}}\mathcal{R}\{d_j\}, \forall
j\in K\\
\label{kkt4}
&\hspace{-1.2cm}\frac{d\mathcal{L}(\mathbf{x},\mu_j,\alpha_j,\lambda)}{d\lambda}&=\bigg(\mathbf{x}^H\mathbf{h}^H_{ps}\mathbf{h}_{ps}\mathbf{x}-\mathcal{I}_{th}\bigg)
\end{eqnarray}
By equating $\frac{d\mathcal{L}(\mathbf{x},\mu_j,\alpha_j,\lambda)}{d\mathbf{x}^{*}}$
to zero, we can formulate $\mathbf{x}$ as the following expression
\begin{eqnarray}
\label{x}
\mathbf{x}=(\mathbf{I}+\lambda\mathbf{h}^H_{ps}\mathbf{h}_{ps})^{-1}\bigg(\sum_j0.5i\mu_j\mathbf{h}^H_{ss,j}+\sum_j 0.5\alpha_j\mathbf{h}^H_{ss,j}\Bigg).
\end{eqnarray}
By substituting (\ref{x}) in the set of (\ref{kkt4})
to form the set of equations (\ref{setoo}),
we can find the solution of $\lambda$, $\alpha_j$ and $\mu_j$ that satisfies
the constraints.
\begin{figure*}[t]
\vspace{-0.2cm}
\hspace{0.2cm}
\begin{tabular}[t]{c}
\begin{minipage}{15 cm}
\scriptsize
 \begin{eqnarray}
\label{multicasteq}
\begin{array}{cccc}
\label{setoo}
\hspace{-4.5cm}\bigg(\sum_j0.5i\mu^*_j\mathbf{h}_{ss,j}+\sum_j 0.5\alpha^*_j\mathbf{h}_{ss,j}\Bigg)(\mathbf{I}+\lambda\mathbf{h}^H_{ps}\mathbf{h}_{ps})^{-1}&\mathbf{h}^H_{ps}\mathbf{h}_{ps}&(\mathbf{I}+\lambda\mathbf{h}^H_{ps}\mathbf{h}_{ps})^{-1}\bigg(\sum_j0.5i\mu_j\mathbf{h}^H_{ss,j}+\sum_j 0.5\alpha_j\mathbf{h}^H_{ss,j}\Bigg)\leq\mathcal{I}_{th}\\
\hspace{-4.5cm}-0.5i(\mathbf{h}_{ss,1}(\mathbf{I}+\lambda\mathbf{h}^H_{ps}\mathbf{h}^H_{ps})^{-1}\bigg(\sum_j0.5i\mu_j\mathbf{h}^H_{ss,j}+\sum_j 0.5\alpha_j\mathbf{h}^H_{ss,j}\Bigg)&+&0.5i(\sum_j\mu_j\mathbf{h}_{ss,j}+\sum_j 0.5\alpha_j\mathbf{h}_{ss,j})(\mathbf{I}+\lambda\mathbf{h}^H_{ps}\mathbf{h}^H_{ps})^{-1}\mathbf{h}_{ss,1}=\sqrt{\psi_1\zeta_1}\mathcal{I}\{d_1\}\\
\hspace{-4.5cm}0.5(\mathbf{h}_{ss,j}(\mathbf{I}+\lambda\mathbf{h}^H_{ps}\mathbf{h}^H_{ps})^{-1}\bigg(\sum_j0.5i\mu_j\mathbf{h}^H_{ss,j}+\sum_j 0.5\alpha_j\mathbf{h}^H_{ss,j}\Bigg)&+&0.5(\sum_j\mu_j\mathbf{h}_{ss,j}+\sum_j 0.5\alpha_j\mathbf{h}_{ss,j})(\mathbf{I}+\lambda\mathbf{h}^H_{ps}\mathbf{h}^H_{ps})^{-1}\mathbf{h}_{ss,j}=\sqrt{\psi_1\zeta_1}\mathcal{R}\{d_1\}\\
&\vdots&\\
\hspace{-4.5cm}-0.5i(\mathbf{h}_{ss,K}(\mathbf{I}+\lambda\mathbf{h}^H_{ps}\mathbf{h}^H_{ps})^{-1}\bigg(\sum_j0.5i\mu_j\mathbf{h}^H_{ss,j}+\sum_j 0.5\alpha_j\mathbf{h}^H_{ss,j}\Bigg)&+&0.5i(\sum_j\mu_j\mathbf{h}_j+\sum_j 0.5\alpha_j\mathbf{h}_{ss,j})(\mathbf{I}+\lambda\mathbf{h}^H_{ps}\mathbf{h}^H_{ps})^{-1}\mathbf{h}_{ss,K}=\sqrt{\psi_K\zeta_K}\mathcal{I}\{d_K\}\\
\hspace{-4.5cm}0.5(\mathbf{h}_{ss,K}(\mathbf{I}+\lambda\mathbf{h}^H_{ps}\mathbf{h}^H_{ps})^{-1}\bigg(\sum_j0.5i\mu_j\mathbf{h}^H_{ss,j}+\sum_j 0.5\alpha_j\mathbf{h}^H_{ss,j}\Bigg)&+&0.5(\sum_j\mu_j\mathbf{h}_{ss,j}+\sum_j 0.5\alpha_j\mathbf{h}_{ss,j})(\mathbf{I}+\lambda\mathbf{h}^H_{ps}\mathbf{h}^H_{ps})^{-1}\mathbf{h}_{ss,K}=\sqrt{\psi_K\zeta_1}\mathcal{R}\{d_K\}\\
\end{array}
\end{eqnarray}
\end{minipage}\\
\hline
\hline
\end{tabular}
\normalsize
\end{figure*}

\subsection{Zero Interference Constraint}
If the PU cannot handle any interference,  the cognitive transmissions should
be in the null space of the channel between CBS and PU. The null space can be
defined as
 
\begin{eqnarray}
\mathbf{\Pi}_{\perp\mathbf{h}_{ps}}=\mathbf{I}-\frac{\mathbf{h}^H_{ps}\mathbf{h}_{ps}}{\|\mathbf{h}_{ps}\|^2}.
\end{eqnarray}
We design the output vector $\mathbf{x}$ to span the null space of $\mathbf{h}_{ps}$
as the following
\begin{eqnarray}
\mathbf{{x}}=\mathbf{\Pi}_{\perp\mathbf{h}_{ps}}\mathbf{\hat{x}}.
\end{eqnarray}
 \begin{eqnarray}\nonumber
&\arg&\mathop {\min} \limits {\mathbf{w}_1,\mathbf{w}_2,...,\mathbf{w}_K}\quad \|\sum^K_{k=1}\mathbf{w}_kd_k\|^2\\\nonumber
 \hspace{-0.1cm}s.t.&\mathcal{C}1:&\angle(\mathbf{h}_j\sum^K_{k=1}\mathbf{w}_k
 d_k)=\angle(d_j),
 \forall j\in K\\\nonumber
&\mathcal{C}2:&\|\mathbf{h}_j\sum^K_{k=1}\mathbf{w}_kd_k\|^2\geq\sigma^2\zeta_j\quad
 , \forall j\in K\\\nonumber
&\mathcal{C}_3:&\|\mathbf{h}_{ps}\sum_{k=1}\mathbf{w}_kd_k\|^2 = 0.
\end{eqnarray}\\
The previous optimization can be written as
The Lagrange function of this optimization problem
\begin{eqnarray}
\mathcal{L}(\mathbf{\hat{x}})&=&\|\mathbf{\hat{x}}\|^2\\\nonumber
&+&\sum_j{\hat{\mu}_j}\bigg(-0.5i\small(\mathbf{h}_{ss,j}\mathbf{\hat{x}}-\mathbf{\hat{x}}^H\mathbf{h}^H_{ss,j}\small)-\sqrt{\psi_j\zeta_{j}}\mathcal{I}\{d_j\}\bigg)\\\nonumber
&+&\sum_j{\hat{\alpha}_j}\bigg(0.5\small(\mathbf{h}_{ss,j}\mathbf{\hat{x}}+\mathbf{\hat{x}}^H\mathbf{h}^H_{ss,j}\small)
-\sqrt{\psi_j\zeta_{j}}\mathcal{R}\{d_j\}\bigg).
\end{eqnarray}
The KKT condition can be written as
\begin{eqnarray}\nonumber
&\hspace{-0.5cm}\frac{d\mathcal{L}(\mathbf{x},\mu_j,\alpha_j,\lambda)}{d\mathbf{\hat{x}}^*}&=\mathbf{\hat{x}}+\sum
0.5i\mu_j\mathbf{h}^H_{ss,j}+\sum_j 0.5\alpha_j\mathbf{h}^H_{ss,j}+\lambda\mathbf{h}^H_{ps}\mathbf{h}_{ps}\mathbf{\hat{x}}\\\nonumber
\label{kkt2}
&\hspace{-0.5cm}\frac{d\mathcal{L}(\mathbf{x},\mu_j,\alpha_j,\lambda)}{d\mu_j}&=-0.5i\small(\mathbf{h}_{ss,j}\mathbf{\hat{x}}-\mathbf{\hat{x}}^H\mathbf{h}^H_{ss,j}\small)-\sqrt{\psi_j\zeta_{j}}\mathcal{I}\{d_j\},
\forall
j\in K\\\nonumber
\label{kkt3}
&\hspace{-0.5cm}\frac{d\mathcal{L}(\mathbf{x},\mu_j,\alpha_j,\lambda)}{d\alpha_j}&=0.5\small(\mathbf{h}_{ss,j}\mathbf{\hat{x}}+\mathbf{\hat{x}}^H\mathbf{h}^H_{ss,j}\small)
-\sqrt{\psi_j\zeta_{j}}\mathcal{R}\{d_j\}, \forall
j\in K\\
\end{eqnarray}
The solution for the previous optimization problem can be written as
\begin{eqnarray}
\mathbf{\hat{x}}=\sum_j0.5i\hat{\mu}_j\mathbf{h}^H_{ss,j}+\sum_j 0.5\hat{\alpha}_j\mathbf{h}^H_{ss,j}
\end{eqnarray}
where $\hat{\mu}_j$, $\hat{\alpha}_j$ can be found by solving the set of
equation (\ref{multicasteq}). Hence, the final formulation for the solution
at zero interference temperature constraint
\begin{eqnarray}
\mathbf{x}=\bigg(\mathbf{I}-\frac{\mathbf{h}^H_{ps}\mathbf{h}_{ps}}{\|\mathbf{h}_{ps}\|^2}\bigg)\bigg(\sum_j0.5i\hat{\mu}_j\mathbf{h}^H_{ss,j}+\sum_j 0.5\hat{\alpha}_j\mathbf{h}^H_{ss,j}\bigg)
\end{eqnarray}
\begin{figure*}[t]
\vspace{-0.2cm}
\hspace{0.2cm}
\begin{tabular}[t]{c}
\begin{minipage}{15 cm}
 \begin{eqnarray}
\label{multicasteq}
\begin{array}{cccc}
0.5K\|\mathbf{h}_{s1}\|(\sum_k(-\mu_k+\alpha_ki)\|\mathbf{h}_{ss,k}\|\rho_{1k}&-&\sum_k(-\mu_k+\alpha_ki)\|\mathbf{h}_{ss,k}\|\rho^{*}_{1k})=\sqrt{\psi_1\zeta^{}_{1}}\mathcal{I}(d_1)\\
0.5K\|\mathbf{h}_{ss,1}\|(\sum_k(-\mu_ki-\alpha_k)\|\mathbf{h}_{ss,k}\|\rho_{1k}&+&\sum_k(-\mu_ki-\alpha_k)\|\mathbf{h}_{ss,k}\|\rho^{*}_{1k})=\sqrt{\psi_1\zeta^{}_{1}}\mathcal{R}(d_1)\\
\quad&\vdots&\\
0.5K\|\mathbf{h}_{ss,K}\|(\sum_k(-\mu_k+\alpha_ki)\|\mathbf{h}_{ss,k}\|\rho_{Kk}&-&\sum_k(-\mu_k+\alpha_ki)\|\mathbf{h}_{ss,k}\|\rho^{*}_{Kk})=\sqrt{\psi_K\zeta_{K}}\mathcal{I}(d_K)\\
0.5K\|\mathbf{h}_{ss,K}\|(\sum_k(-\mu_ki-\alpha_k)\|\mathbf{h}_{ss,k}\|\rho_{Kk}&+&\sum_k(-\mu_ki-\alpha_k)\|\mathbf{h}_{ss,K}\|\rho^{*}_{Kk})=\sqrt{\psi_K\zeta_{K}}\mathcal{R}(d_K)\\
\end{array}
\end{eqnarray}
\end{minipage}\\
\hline
\hline
\end{tabular}
\end{figure*}
\vspace{-0.2cm}
\section{Theoretical upper-bound}
The theoretical upper-bound can be formulated by dropping the phase constraint
$\mathcal{C}_1$
of (\ref{pow}). The optimal input covariance $\mathbf{Q}$ can be found by solving the following optimization problem:
  
\begin{eqnarray}\nonumber
\mathbf{Q}&=&\arg\mathop {\min}\limits_{\mathbf{Q}}\quad tr(\mathbf{Q})\\\nonumber
&s.t.&\mathbf{h}_{ss,j}\mathbf{Q}\mathbf{h}^H_{ss,j}=\psi_j\gamma_j\forall j\in K.\\
&\quad&\mathbf{h}_{ps}\mathbf{Q}\mathbf{h}^H_{ps}\leq \mathcal{I}_{th}.
\end{eqnarray}
where $\mathbf{Q}=\mathbf{x}\mathbf{x}^H$. This problem resembles the multicast
problem \cite{multicast} with additional interference temperature constraint
to suit the constraint imposed by the primary system.
\vspace{-0.2cm}
\section{Numerical results}
In order to assess the performance of the proposed transmissions schemes, Monte-Carlo simulations of the different algorithms have been conducted to
study the performance of the proposed techniques and compare to the state of the art techniques. The adopted channel model is assumed
to be as the following
\begin{itemize}
\item $\mathbf{h}_{pp}\sim \mathcal{CN}(0,\sigma^2_{pp}\mathbf{1}_{1\times M})$, where $\mathbf{1}_{1\times M}$ is vector of all ones and of size $1\times M$.
\item $\mathbf{h}_{ps}\sim\mathcal{CN}(0,\sigma^2_{ps}\mathbf{1}_{1\times M})$
\item $\mathbf{h}_{ss,j}\sim\mathcal{CN}(0,\sigma^2_{ss}\mathbf{1}_{1\times M}),\forall j\in K$
\item $\mathbf{h}_{sp,j}\sim\mathcal{CN}(0,\sigma^2_{sp,j}\mathbf{1}_{1\times M}),\forall j\in K$
\item To study the performance of the system at the worst case scenario, when all users have a strong channel with respect to its direct and interfering base stations $\sigma^2_{pp}=\sigma^2_{ps}=\sigma^2_{sp,j}=\sigma^2_{ss}=\sigma^2$.
\end{itemize}
In the figures, we denote the proposed cognitive technique that exploits
the constructive interference by (CCIPM), while S denotes the strict interference
constraints $\mathcal{I}_{th}=0$. CBS has 3 antennas and serves 2 cognitive users. We compare the performance of the proposed scheme (CCIPM) to the scheme in \cite{Christos} tailored to cognitive scenario by solving the following optimization
\begin{eqnarray}\nonumber
\mathbf{W}_{CCIZF}=&\mathop {\min}\limits_{\mathbf{W}}& \mathbb{E}\{\|\mathbf{R}_{\phi}-\mathbf{H}_{ss}\mathbf{W}\mathbf{d}\|^2\}\\\nonumber
&s.t.&\|\mathbf{W}\|^2\leq P\\
&\quad&\mathbf{h}_{ps}\mathbf{W}\mathbf{W}^H\mathbf{h}^H_{ps}=0
\end{eqnarray}
where $\mathbf{R}_{\phi}$ is defined in \cite{Christos}. We utilize the energy efficiency metric to assess the performance of the proposed technique as  
\begin{eqnarray}
\eta=\frac{\sum^K_{j=1} R_j}{\|\mathbf{x}\|^2},
\end{eqnarray}
$R_j$ denotes the rate of the $j^{th}$ user. In all figures, we depict the performance of constructive interference zero forcing (CCIZF).
For the sake of comparison, the transmit power of the CCIZF solutions can be scaled until all users achieve the target rate.\smallskip

In Fig.\ref{fig1}, the energy efficiency with respect to the average channel is depicted, the used modulation is QPSK and the strict interference constraint imposed by the primary system. We compare the performance of the
CCIPM to the theoretical upper-bound and CCIZF. It can be noted that the
CCIZF curve saturates at the low-mid SNR regime, while the curves of CCIPM and multicast have higher growth in terms of energy efficiency in the same regime. Moreover, it can be noted that CCIPM outperforms CCIZF at different channel strength values.  

In Fig. \ref{fig2}, we depict the energy efficiency at different target rates. For the constructive interference schemes, we assign the target rate
with its corresponding MPSK modulation. It can be noted that the theoretical
upper bounds for the scenario of the strict and the relaxed interference
constraints have the same power consumption at target rate equals to 1 bps$/$Hz. However, this result does not hold for constructive interference
technique. Moreover, it can be noted that the gap between the theoretical
bound and the CCIPM is fixed at the both scenario for all target rates. 
  
\begin{figure}
\includegraphics[scale=0.65]{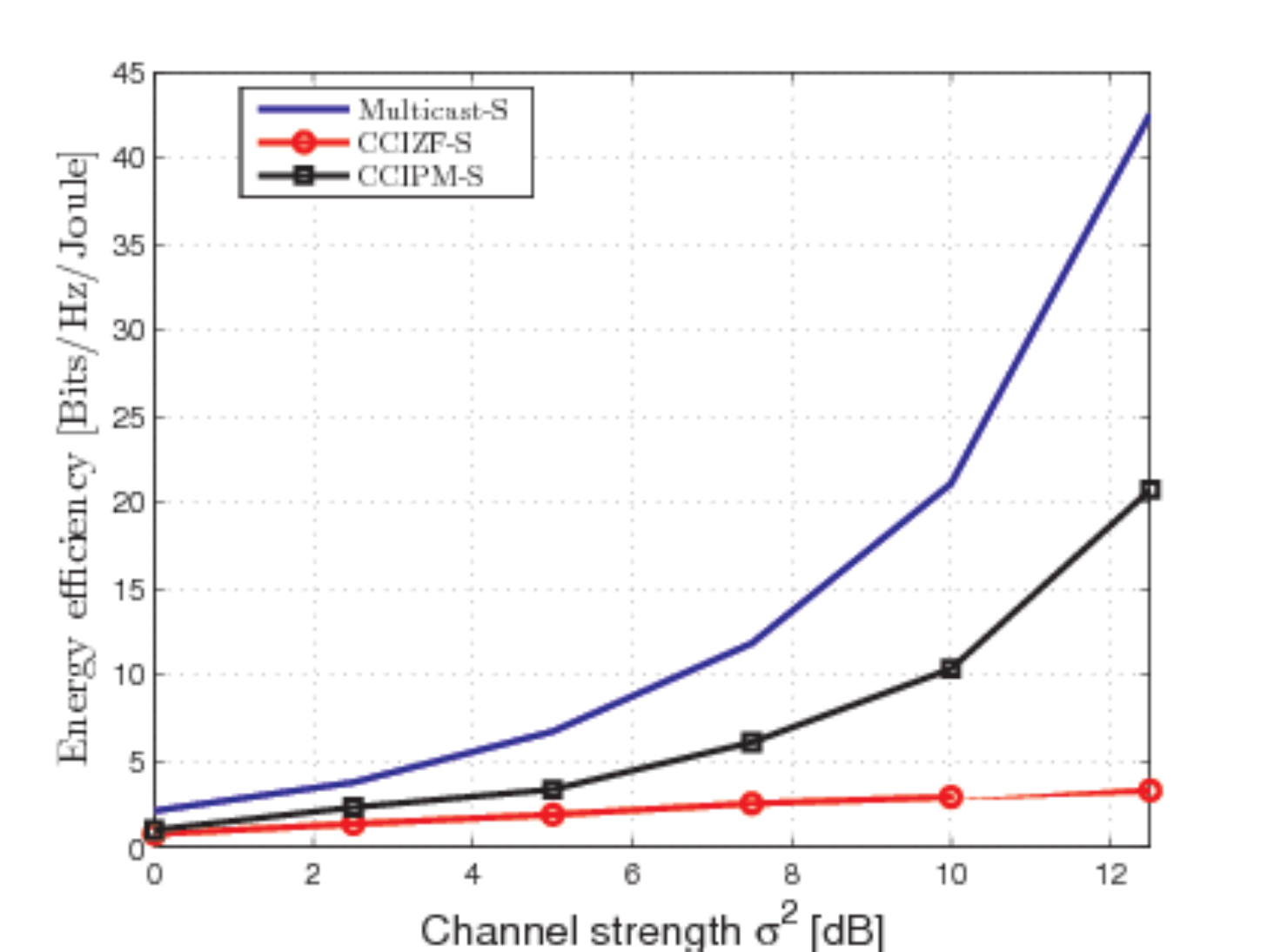}
\caption{\label{ci}\textit{\label{fig1}\small Energy efficiency
vs channel strength. The adopted modulation is QPSK for CCIZF and CCIPM, $\zeta=4.7712dB$. }}
\end{figure}
\begin{figure}
\includegraphics[scale=0.65]{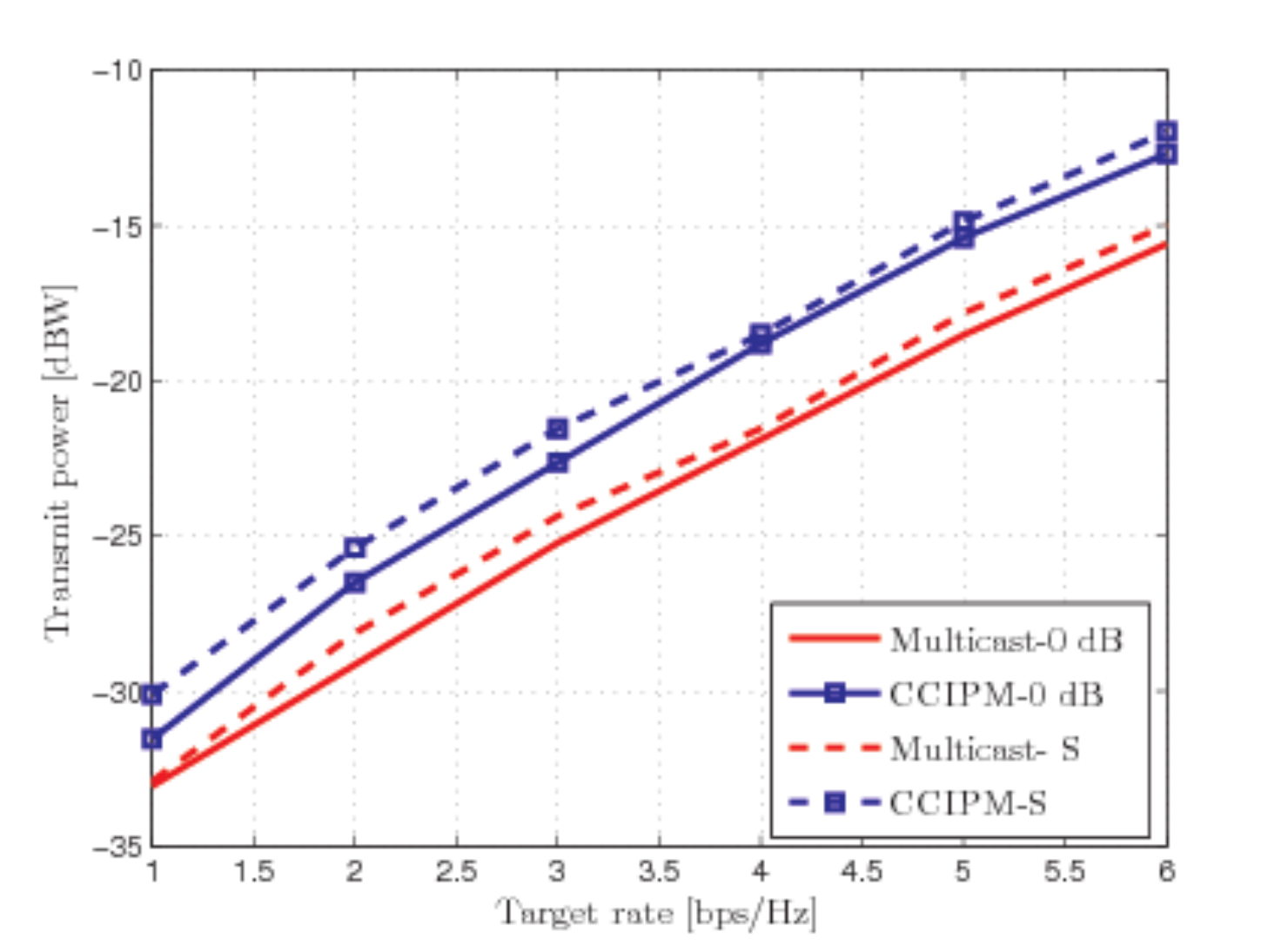}
\caption{\label{ci}\textit{\label{fig2}\small Transmit power vs target
rate. $\sigma^2=10dB$}}
\end{figure}
\section{Conclusions}
In this paper, we propose symbol-level precoding techniques for the downlink of cognitive underlay system. These techniques exploits the availability of the CSI and the data symbols to  constructively correlate the transmission for the cognitive users without violating the interference temperature at the primary users. This enables interference exploitation among the cognitive multiuser transmissions assuming M-PSK modulation. The designed precoder aims at minimizing the transmitted power at CBS while granting a certain received SNR at each cognitive users. From the numerical results section, it can be concluded that the CBS consumes less power at the relaxed interference constraints. Finally, a comparison with the theoretical upperbound and the state-of-the art techniques is illustrated.


\vspace{-0.2cm}
\begin{thebibliography}{1}
\vspace{-0.2cm}
\bibitem{federal}
Federal Communication Commission, Spectrum Policy Task Force, ET
document no. 02-135, Nov. 2002.

\bibitem{goldsmith}
A. Goldsmith, S. A. Jafar, I. Maric, and S. Srinivasa, ``Breaking spectrum
gridlock with cognitive radios: An information theoretic perspective,"
\textit{IEEE}, vol. 97, no. 5, pp. 894 - 914, May 2009.
\bibitem{Haykin}
S. Haykin, ``Cognitive Radio: Brain-Empowered Wireless Communications,"
\textit{IEEE Journal on Selected Areas in Communications}, vol. 23, pp. 201-22, Feb. 2005.
\bibitem{shree}
S. Srinivasa, S. A. Jafar, ``Soft Sensing and Optimal Power Control for Cognitive Radio,"\textit{IEEE
Transactions on Wireless Communications}, vol. 9, no. 12, pp. 3638-3649,
October  2010.
\bibitem{lv_o}
J. Lv, R. Blasco-Serrano, E. Jorswieck, R. Thobaben and A. Kliks,
``Optimal Beamforming in MISO Cognitive Channels with Degraded Message Sets," \textit{IEEE Conference on Wireless Communications and Networking (WCNC)}, April 2012.
\bibitem{jorswieck}
J. Lv, R. Blasco-Serrano, E. Jorswieck and R. Thobaben, ``Linear Precoding
in MISO Cognitive Channels with Causal Primary Message," \textit{IEEE International
Symposium on Wireless Communications Systems (ISWCS)}, 2012.\smallskip
\bibitem{learning}
J. Lv, R. Blasco-Serrano, E. Jorswieck and R. Thobaben,``Multi-antenna transmission for underlay and
overlay cognitive radio with explicit
message-learning phase,"EURASIP Journal on Wireless Communications and Networking (JWCN), special issue on "Cooperative Cognitive Networks", Jul. 2013. 
\bibitem{gan_t}
G. Zheng, S. H. Song, K. K. Wong, and B. Ottersten,``Cooperative Cognitive Networks: Optimal, Distributed and Low-Complexity Algorithms," \textit{IEEE Transaction on Signal Processing}, vol. 61, no.11 , pp. 2778 - 2790, June 2013.
\bibitem{ben}
S. H. Song and K.B. Letaief,``Prior Zero-Forcing for Relaying Primary Signals in Cognitive Network," \textit{IEEE Global Telecommunication Conference (GLOBECOM)}, December 2010. 
\bibitem{multicast}
N. D. Sidropoulos, T. N. Davidson, ans Z.-Q. Luo, ``Transmit Beamforming
for Physical-Layer Multicasting," \textit{IEEE Transactions on Signal Processing},
vol. 54, no. 6, pp. 2239-2251, June 2006.

 \bibitem{multicast-jindal}
 N. Jindal and Z.-Q. Luo, ``Capacity Limits of Multiple Antenna Multicast,"
\textit{IEEE International Symposium on Information Theory (ISIT)}, pp.
 1841 - 1845, June 2006. 

\bibitem{Christos-1}
C. Masouros and E. Alsusa,``Dynamic Linear Precoding for the exploitation of Known Interference in MIMO Broadcast Systems," \textit{IEEE Transactions On Communications}, vol. 8,
no. 3, pp. 1396 - 1404, March 2009.
\bibitem{Christos}
C. Masouros, ``Correlation Rotation Linear Precoding for MIMO Broadcast Communications,"
\textit{IEEE Transaction on Signal Processing}, vol. 59, no. 1, pp. 252 -262, January 2011.
\bibitem{isspit}
M. Alodeh, S. Chatzinotas and B. Ottersten, ``Data Aware User Selection
in the Cognitive Downlink MISO Precoding Systems," \textit{invited paper
to IEEE International
Symposium on Signal Processing and Information Technology (ISSPIT)}, December
2013.
\bibitem{maha}
M. Alodeh, S. Chatzinotas and B. Ottersten, ``A Multicast Approach for Constructive Interference Precoding in MISO Downlink Channel,"\textit{  in the proceedings
 of International Symposium in Information theory (ISIT)
2014, Available on arXiv:1401.6580v2 [cs.IT].}
\bibitem{maha_TSP}
M. Alodeh, S. Chatzinotas and B. Ottersten, ``Constructive Multiuser Interference in Symbol Level Precoding for the MISO Downlink Channel,"\textit{ IEEE Transactions on Signal processing
2015, Available on arXiv:1408.4700 [cs.IT].}

\end{thebibliography}
\end{document}